\documentclass{article}
\usepackage{spconf,amsmath,graphicx,hyperref}

\usepackage{cite}
\usepackage{amssymb,amsfonts}
\usepackage{algorithm}
\usepackage{booktabs} 
\usepackage{array}    
\usepackage{algorithmic}
\usepackage{textcomp}
\usepackage{xcolor}
\usepackage{cite}
\usepackage{subcaption}
\def\BibTeX{{\rm B\kern-.05em{\sc i\kern-.025em b}\kern-.08em
    T\kern-.1667em\lower.7ex\hbox{E}\kern-.125emX}}
    
\newtheorem{prop}{Proposition}[section]

\newcommand{\Toep}{\mathrm{Toep}}

\newcommand{\complex}{\mathbb{C}}

\newcommand{\cA}{\mathcal{A}}

\newcommand{\cB}{\mathcal{B}}

\newcommand{\bB}{{\boldsymbol{B}}}

\newcommand{\bd}{{\boldsymbol{d}}}

\newcommand{\bD}{\boldsymbol{D}}
\newcommand{\bn}{\boldsymbol{n}}

\newcommand{\bT}{\boldsymbol{T}}

\newcommand{\bx}{\boldsymbol{x}}

\newcommand{\bA}{\boldsymbol{A}}
\newcommand{\by}{\boldsymbol{y}}
\newcommand{\bY}{\boldsymbol{Y}}
\newcommand{\bX}{\boldsymbol{X}}

\newcommand{\bZ}{\boldsymbol{Z}}

\newcommand{\be}{\boldsymbol{e}}
\newcommand{\ba}{\boldsymbol{a}}
\newcommand{\bb}{\boldsymbol{b}}

\newcommand{\bg}{\boldsymbol{g}}
\newcommand{\bh}{\boldsymbol{h}}

\newcommand{\bz}{\boldsymbol{z}}

\newcommand{\bu}{\boldsymbol{u}}

\newcommand{\bI}{\boldsymbol{I}}

\newcommand{\norm}[2][2]{\left\| #2 \right\|_{#1}}

\title{A Convex Demixing Approach for Hybrid-Field Channel Estimation \\ of XL-MIMO Systems via Atomic Norm Minimization}
\twoauthors
  {Dehui Yang}
	{School of Mathematics and Physics \\ Xi’an Jiaotong-Liverpool University\\ Suzhou 215123, China}
  {Feng Xi, Yanxian Zhu}
	{Department of Electronic Engineering \\
	Nanjing University of Science and Technology \\
	Nanjing 210094, China}
	
\begin{document}
%
\maketitle
\begin{abstract}
Channel estimation is a critical task in extremely large-scale multiple-input multiple-output (XL-MIMO) systems for 6G wireless communications. A hybrid-field channel model effectively characterizes the mixed far-field and near-field scattering components in practical XL-MIMO systems. In this paper, we propose a convex demixing approach for hybrid-field channel estimation within the atomic norm minimization (ANM) framework. By promoting sparsity of the far-field and near-field components directly in the continuous parameter domain, a demixing scheme that minimizes a weighted sum of two atomic norms is proposed.
We show that the resulting ANM is equivalent to a computationally feasible semi-definite programming (SDP). Numerical experiments on simulated data demonstrate that our method outperforms existing approaches for hybrid-field channel estimation.  
\end{abstract}
\section{Introduction}
\label{sec:intro}
Extremely large-scale MIMO has emerged as a key technology for enhancing spectral efficiency in future 6G wireless communication systems  \cite{Saad-MNET2020,Wang-COMST2024}. 
To achieve high spectral and energy efficiency and support high data rates, a large number of antennas are typically employed at the base station (BS). As a result, channel estimation becomes a challenging task due to a prohibitively high pilot overhead. By exploiting the sparse representation of the channel, two reduced-overhead channel estimation schemes have been developed, based on the Rayleigh distance separating the electromagnetic (EM) radiation field into far-field and near-field regions \cite{Cui-ToC2022}. In the far-field region, the phase of an EM wave is well-approximated by a first-order Taylor expansion, resulting in negligible
phase discrepancy. Under the planar wave assumption, channel sparsity in the angular domain has been widely leveraged, and numerous sparse recovery algorithms have been developed \cite{Cui-ToC2022,Lee-TCOMM2016,Huang-TSP2019}. In contrast, in the near-field region, the spherical wave assumption supports a polar-domain sparse representation, with steering vectors parametrized by both angle and range. Techniques, including discretized sparse recovery \cite{Pisharody-TWC2024,Zhang-2024ToC,near-field-dai-tutorial} and gridless atomic norm minimization methods \cite{Li-ICASSP2025}, have been proposed for near-field channel estimation. 

In practical XL-MIMO systems, scatters are often distributed across both far-field and near-field regions. For example, a scatterer close to the BS creates a near-field path, while a direct link from a user far away from the BS forms a far-field path. Thus, a hybrid-field channel model, capturing both far-field and near-field components, is more practical and accurate \cite{Wei-CL2022,near-field-tutorial}. Recently, sparse recovery approaches have been devised to estimate the far-field and near-field components of the hybrid-field channel model \cite{Wei-CL2022,Lei-TSP2024}. However, these approaches require discretization and estimation of the number of far-field and near-field components,  introducing basis mismatch and reducing channel estimation accuracy. 

In this paper, we propose a convex demixing scheme for efficient estimation of far-field and near-field channel components as well as the associated parameters from their mixtures. Specifically, we represent each near-field component's steering vector as a far-field steering vector that depends only on the unknown angle parameter, modulated by an unknown waveform. By representing the unknown modulation waveforms using an effective subspace model, we reformulate the problem as minimization of a weighted combination of two atomic norms that promote sparsity for the far-field and near-field components without discretizing the angle parameter. The resulting ANM has an equivalent SDP, which is solvable using off-the-shelf convex optimization software toolboxes. Numerical simulations demonstrate superior channel estimation accuracy is achieved by the proposed ANM, compared to existing discretized sparse recovery methods. 


\section{System and Channel Models}
\subsection{System Model}
\label{subsec:Model}
Consider an uplink multiuser XL-MIMO system serving $U$ single-antenna users. The BS employs a hybrid analog and digital transceiver architecture to reduce the number of required radio frequency (RF) chains.
Specifically, the BS uses a uniform linear array (ULA) with $N$ antennas and $N_{\text{RF}}$ RF chains, where $N_{\text{RF}}\ll N$. Each RF chain is connected to the antennas through analog phase shifters. The antenna elements are spaced by $d=\lambda_c/2$, where $\lambda_c$ denotes the carrier wavelength.
The antennas are arranged along the $y$-axis, with the $n$th antenna located at position $(0,(n-1)d)$, for $n=1,\cdots,N$. 
During uplink channel estimation, $U$ users transmit mutually orthogonal pilot sequences to the BS, where $U\leq N_{\text{RF}}$.
This setup enables independent channel estimation for each user, as discussed in \cite{Emil-2016CM}. 

Without loss of generality, we consider a user located at $(x, y)$ with polar coordinates $(r,\theta) = (\sqrt{x^2+y^2}, \arctan \frac{y}{x})$. Each user transmits a pilot sequence $\{s_1, s_2,\cdots, s_P\}$ of length $P$ during the channel estimation phase, where $s_{p}$ denotes the pilot signal transmitted in the $p$th time slot. 
The received pilot signal in the $p$th time slot is given by
\begin{equation}
    \tilde{\by}_{p} = \bA_{p}(\bh s_{p} + \bn_{p})\in\mathbb{C}^{N_{\text{RF}}\times 1},
\end{equation}
where $\bA_{p}\in\mathbb{C}^{N_{\text{RF}}\times N}$ is the analog combining matrix, $\bh\in\mathbb{C}^{N\times 1}$ represents the channel vector between the BS and each user, and $\bn_{p}\in\mathbb{C}^{N\times 1}$ denotes the additive Gaussian noise vector, distributed as $\bn_{p}\sim\mathcal{CN}(0,\sigma^2 \mathbf{I}_N)$, where $\sigma^2$ is the noise variance.
The analog combining matrix $\bA_{p}$ is typically implemented using analog phase shifters and satisfies the constant modulus constraint, i.e., $|[\bA_{p}]_{i,j}| = \frac{1}{\sqrt{N}}$.
Define $\by_{p} = \frac{\overline{s}_{p}}{|s_{p}|^2}\tilde{\by}_{p}$ and $\by=[\by_1^{T},\cdots,\by_P^{T}]^{T}\in\mathbb{C}^{N_{\text{RF}}P\times 1}$, where $\overline{\cdot}$ denotes the conjugate operation.
After receiving the complete pilot sequence, we have
\begin{equation}\label{eqn:y}
    \by = \bA \bh + \bn,
\end{equation}
where the measurement $\bA = [\bA_1^{T},\cdots,\bA_P^{T}]^{T}\in\mathbb{C}^{N_{\text{RF}}P\times N}$ and $\bn=
[(\frac{\overline{s}_{p}}{|s_{p}|^2} \bA_1\bn_1)^{T},\cdots,(\frac{\overline{s}_{p}}{|s_{p}|^2} \bA_P\bn_P)^{T}]^{T}\in\mathbb{C}^{N_{\text{RF}}P\times 1}$.

In XL-MIMO systems, the number of antennas $N$ at the BS is often significantly larger than the number of measurements $N_{\text{RF}}P$ at the receiver, rendering channel estimation ill-posed.
To address this challenge, exploiting the inherent structure of the channel vector $\bh$ is crucial for effective channel estimation.

\subsection{Hybrid-Field Channel Model}
Consider the following hybrid-field channel model 
\begin{equation}
\label{eq:hybrid-field}
\bh_{\text{hybrid}} = \underbrace{\sqrt{\frac{N}{K}} \sum_{k_\text{f} = 1}^{K_\text{f}} \alpha_{k_\text{f}} \ba(\theta_{k_\text{f}})}_{\bh_{\text {far}}} + \underbrace{\sqrt{\frac{N}{K}} \sum_{k_\text{n} = 1}^{K_\text{n}} \alpha_{k_\text{n}} \bb(\theta_{k_\text{n}}, r_{k_\text{n}})}_{\bh_\text{near}},
\end{equation}
where $K_\text{f}$ and $K_{\text{n}}$ represent the numbers of far-field and near-field paths, respectively, $K = K_\text{f} + K_\text{n}$ is the total number of propagation paths, and $\alpha_{k_\text{f}}$ and $\alpha_{k_\text{n}}$ are the complex gains for the far-field and near-field paths, respectively. The far-field and near-field steering vectors, $\ba(\theta_{k_\text{f}}) \in \mathbb{C}^N$ and $\bb(\theta_{k_\text{n}}, r_{k_\text{n}}) \in \mathbb{C}^N$, are defined as
\begin{equation*}
\begin{cases}
 \ba(\theta_{k_\text{f}})  = \begin{bmatrix}1, e^{j\frac{2\pi}{\lambda_c} d \sin \theta_{k_\text{f}}}, \cdots, e^{j\frac{2\pi}{\lambda_c} d(N-1)\sin \theta_{k_\text{f}}}\end{bmatrix}^T \\
 \bb(\theta_{k_\text{n}}, r_{k_\text{n}}) = \begin{bmatrix} e^{-j \frac{2\pi}{\lambda_c} (r_{k_\text{n}}^{(1)} - r_{k_\text{n}})}, \cdots, e^{-j \frac{2\pi}{\lambda_c} (r_{k_\text{n}}^{(N)} - r_{k_\text{n}})} \end{bmatrix}^T,
\end{cases}
\end{equation*} 
where $\theta_{k_\text{f}}$ and $\theta_{k_\text{n}}$ are the continuous angle parameters,  $r_{k_\text{n}}$ is the range from the $k$th scatterer to the reference antenna, and $r_{k_\text{n}}^{(n)}$ denotes the range from the $k$th scatterer to the $n$th antenna at position $(0, (n-1)d)$. By geometry, it is easy to see that
$ r_{k_\text{n}}^{(n)} = \sqrt{r_{k_\text{n}}^2 + ((n-1)d)^2 - 2r_{k_\text{n}} (n-1)d \sin \theta_{k_\text{n}}}.
$

The hybrid-field channel model (\ref{eq:hybrid-field}) generalizes the far-field and near-field models: when $K_\text{f} = K$ and $K_\text{n} = 0$, (\ref{eq:hybrid-field}) reduces to the far-field channel model, and when $K_\text{n} = K$ and $K_\text{f} = 0$, it becomes the near-field channel model. Therefore, the hybrid-field channel captures scatters in both regions, providing a more flexible and robust channel model for XL-MIMO systems. However, channel estimation for the hybrid-field model is more complicated, as it requires demixing the far-field and near-field components from their mixtures alongside parameter estimation. 
\label{sec:format}
\section{Methodology}
This section presents a convex program for hybrid-field channel estimation, simultaneously demixing far-field and near-field components. 
\label{sec:pagestyle}
\subsection{Atomic Norm Minimization for Hybrid-Field Channel Estimation}
To start with, observe that the range $r_{k_\text{n}}^{(n)}$ can be approximated using a second-order Taylor expansion:
\begin{equation}
\label{eq:approx}
\begin{aligned}
r_{k_\text{n}}^{(n)} & =  \sqrt{r_{k_\text{n}}^2 + ((n-1)d)^2 - 2r_{k_\text{n}} (n-1)d \sin \theta_{k_\text{n}}} \\
& \approx r_{k_\text{n}} - (n-1)d\sin \theta_{k_\text{n}} + \frac{(n-1)^2 d^2 \cos^2 \theta_{k_\text{n}}}{2r_{k_\text{n}}},
\end{aligned}
\end{equation}
where the approximation follows from the Taylor expansion $\sqrt{(1 + x)} \approx 1 + \frac{1}{2} x - \frac{1}{8} x^2 + o(x^2)$. 
Substituting (\ref{eq:approx}) into the near-field steering vector $ \bb(\theta_{k_\text{n}}, r_{k_\text{n}})$, the $n$th component of $\bb(\theta_{k_\text{n}}, r_{k_\text{n}})$ can be approximated as 
\begin{equation}
\left[\bb(\theta_{k_\text{n}}, r_{k_\text{n}})\right]_n \approx   e^{j \frac{2\pi}{\lambda_c} \left[(n-1) d\sin \theta_{k_\text{n}} - \frac{(n-1)^2 d^2 \cos^2 \theta_{k_\text{n}}}{2r_{k_\text{n}}}\right]}.
\end{equation}
Define the parameters $\phi_{k_\text{f}} = \frac{d}{\lambda_c} \sin \theta_{k_\text{f}}$, $\phi_{k_\text{n}} = \frac{d}{\lambda_c} \sin \theta_{k_\text{n}}$, and $\psi_{k_\text{n}} =  - \frac{d^2 \cos^2\theta_{k_\text{n}}}{2 \lambda_c r_{k_\text{n}}}$. By a change of variables, we can compactly rewrite the hybrid-field channel model (\ref{eq:hybrid-field}) approximately as
\begin{equation}
\label{eq:approx-hybrid-model}
\bh_{\text{hybrid}} \approx \sqrt{\frac{N}{K}}  \left(\sum_{k_\text{f} = 1}^{K_\text{f}} \alpha_{k_\text{f}} \bd (\phi_{k_\text{f}}) + \sum_{k_\text{n} = 1}^{K_\text{n}} \alpha_{k_\text{n}} \bd(\phi_{k_\text{n}})\odot \bg(\psi_{k_\text{n}})\right),
\end{equation}
where $\bd(\phi_{k_f})$, $\bd(\phi_{k_n})$, and $\bg(\psi_{k_n})$ are defined as
\begin{equation*}
\begin{cases}
\bd (\phi_{k_\text{f}})  =  \begin{bmatrix}1, e^{-j2\pi \phi_{k_\text{f}}}, \cdots, e^{-j 2\pi (N-1) \phi_{k_\text{f}}}\end{bmatrix}^H,\\
\bd (\phi_{k_\text{n}})  =  \begin{bmatrix}1, e^{-j2\pi \phi_{k_\text{n}}}, \cdots, e^{-j 2\pi (N-1) \phi_{k_\text{n}}}\end{bmatrix}^H,\\ 
 \bg(\psi_{k_\text{n}}) = \begin{bmatrix} 1,  e^{-j2\pi 1^2 \psi_{k_\text{n}}}, \cdots, e^{-j 2\pi (N-1)^2 \psi_{k_\text{n}}} \end{bmatrix}^H
\end{cases}
\end{equation*}
and $\odot$ denotes the Hadamard product. 
Thus, estimating the original parameters $\left\{(\theta_{k_\text{f}}, \theta_{k_\text{n}}, r_{k_\text{n}})\right\}$ in (\ref{eq:hybrid-field}) is equivalent to recovering the transformed parameters $\left\{(\phi_{k_\text{f}}, \phi_{k_\text{n}}, \psi_{k_\text{n}})\right\}$ in (\ref{eq:approx-hybrid-model}).

The atomic norm, first proposed in \cite{chandrasekaran2012convex},
provides a general framework for enforcing sparsity in a signal or dataset that is a superposition of a few atoms from a dictionary \cite{tang2013compressed,li2014off,Yang-2016TSPMMV}. 
To promote sparsity for the far-field channel component $\bh_{\text{far}}$, define the atomic norm associated with the set
of atoms 
$
\cA_{\text{f}} = \left\{\bd(\phi)~|~\phi \in [0, 1)\right\}
$
as 
$\norm[\cA_\text{f}]{\bx}  
  = \inf_{\alpha_i, \phi_i \in [0, 1)}\left\{\sum_i |\alpha_i|:~\bx = \sum_i \alpha_i \bd(\phi_i) \right\}$.

For the near-field channel component $\bh_{\text{near}}$, where each steering vector $\bd (\phi_{k_\text{n}})$ is modulated by an unknown waveform $ \bg(\psi_{k_\text{n}})$, following the work in \cite{Yang-2016TIT},
we assume $\left\{\bg(\psi_{k_\text{n}})\right\}$ lies in a known $L$-dimensional subspace with $L \ll N$, i.e., 
\begin{equation}
\label{eq:subspace}
\bg(\psi_{k_\text{n}}) = \bB \bz_{k_\text{n}}
\end{equation}
for some matrix $\bB \in \mathbb{C}^{N\times L}$ and unknown coefficient $ \bz_{k_\text{n}} \in \mathbb{C}^{L\times 1}$. 
Methods for constructing $\bB$ include a discretization-based approach, where a dictionary consisting of waveforms $\bg(\psi)$ is built on a fine grid of $ \psi$ followed by principal component analysis (PCA), or the use of Discrete Prolate Spheroidal Sequences (DPSS) \cite{dpss}. Following \cite{Li-ICASSP2025}, we use the discretization based method for constructing the subspace matrix $\bB$. Using the subspace model for $\bg(\psi_{k_\text{n}})$, we have 
\begin{equation}
\label{eq:inner_product}
\begin{aligned}
&  \left[ \sum_{k_\text{n} = 1}^{K_\text{n}} \alpha_{k_\text{n}} \bd(\phi_{k_\text{n}})\odot \bg(\psi_{k_\text{n}})\right]_n \\
 & =  \left\langle \sum_{k_\text{n} = 1}^{K_\text{n}} \alpha_{k_\text{n}}  \bz_{k_\text{n}} \bd(\phi_{k_\text{n}})^H , \bb_n\be_n^H \right\rangle,
\end{aligned}
\end{equation}
where we have defined $\left\langle \bX, \bY \right\rangle = \text{trace}(\bY^H \bX)$ and used the notation 
$\be_n$ to denote the $n$th column of an $N\times N$ identity matrix $\bI_N$. Equation (\ref{eq:inner_product}) indicates that under the subspace model, the $n$th entry of the near-field component can be written as the inner product of a structured low-rank matrix $\bX^{\star} = \sum_{k_\text{n} = 1}^{K_\text{n}} \alpha_{k_\text{n}}  \bz_{k_\text{n}} \bd(\phi_{k_\text{n}})^H$ and the sensing matrix $\bb_n\be_n^H$. Thus, we write 
\begin{equation}
\label{eq:measurement}
\sum_{k_\text{n} = 1}^{K_\text{n}} \alpha_{k_\text{n}} \bd(\phi_{k_\text{n}})\odot \bg(\psi_{k_\text{n}}) = \cB(\bX^{\star}),
\end{equation}
where $\bg(\psi_{k_\text{n}})  = \bB  \bz_{k_\text{n}}$ and the linear operator $\cB: \mathbb{C}^{L\times N} \rightarrow \mathbb{C}^N$ is defined as 
$\left[\cB(\bX^{\star})\right]_n = \left \langle \bX^{\star},  \bb_n\be_n^H\right\rangle$. 
To promote the low-rank structure of $\bX^{\star}$, define the atomic set 
$\cA_\text{n} = \left\{ \bz \bd(\phi)^H:~\phi\in [0,1), \norm{\bz} = 1, \bz \in \complex^{L\times 1} \right\}$
and its atomic norm as 
\begin{equation*}
\begin{aligned}
\norm[\cA_\text{n}]{\bX}  
&  = \inf_{\beta_i, \phi_i, \norm{\bz_i} = 1}\left\{\sum_i |\beta_i|:~\bX = \sum_i \beta_i \bz_i \bd(\phi_i)^H \right\}.
\end{aligned}
\end{equation*}
To enforce sparsity for both the far-field and near-field components using $\norm[\cA_\text{f}]{\bx}$ and $\norm[\cA_\text{n}]{\bX}$, we solve the following convex program with respect to the measurement constraint   
\begin{equation}
\label{eq:atomic norm}
\begin{aligned}
& \mathop{\text{minimize}}\limits_{\bx, \bX}~~\norm[\cA_\text{f}]{\bx} + \tau \norm[\cA_\text{n}]{\bX}  \\
&  {\text{subject~to}}~~\| \by - \bA(\bx + \cB(\bX))\|_2 \leq \delta, 
\end{aligned}
\end{equation}
where the regularization parameter $\tau$ balances the strengths of two atomic norms and $\delta$ bounds the amount of noise. We refer to the convex program (\ref{eq:atomic norm}) as convex demixing for hybrid-field channel estimation. 

Prior work has shown that both $\norm[\cA_\text{f}]{\bx}$ and $\norm[\cA_\text{n}]{\bX}$ admit equivalent SDPs \cite{tang2013compressed,chi2015guaranteed,Yang-2016TIT}. As a result, we have the following proposition.
\begin{prop} The constrained atomic norm minimization (\ref{eq:atomic norm}) is equivalent to the following SDP: 
\begin{equation}
\label{eq:sdp}
\begin{aligned}
& \mathop{\text{minimize}}\limits_{\bx, \bX, \bu_{\text{far}},  \bu_{\text{near}}, t, \bT}~~  \frac{1}{2N} \text{trace}(\text{Toep}(\bu_\text{far})) + \frac{1}{2} t  \\
&~~~~~~~~~~~~~~~~~~~~~~ + \frac{\tau }{2N} \text{trace}(\text{Toep}(\bu_\text{near})) + \frac{\tau }{2} \text{trace}(\bT) \\
& {\text{subject~to}}~~\| \by - \bA(\bx + \cB(\bX))\|_2 \leq \delta \\
& ~~~~\begin{bmatrix} \text{Toep}(\bu_{\text{far}}) & \bx \\ \bx^H & t \end{bmatrix}\succeq 0,~\begin{bmatrix}
\text{Toep}(\bu_\text{near}) & \bX^H \\
\bX & \bT
\end{bmatrix}   \succeq 0,
\end{aligned}
\end{equation}
where $\bu_\text{far}$ and $\bu_\text{near}$ are the complex vectors whose first entries are real, $\Toep(\bu_\text{far})$ and $\Toep(\bu_\text{near})$  denote the $N \times N$ Hermitian Toeplitz matrix whose respective first columns are $\bu_\text{far}$ and $\bu_\text{near}$, and $\bT$ is a Hermitian $L\times L$ matrix.
\end{prop}
After solving the SDP (\ref{eq:sdp}) using packages such as CVX \cite{cvx}, we obtain a solution, denoted as $\{ \widehat{\bx}, \widehat{\bX},  \widehat{\bu}_{\text{far}},  \widehat{\bu}_{\text{near}}, \widehat{t}, \widehat{\bT}\}$. The hybrid-field channel is estimated as $\widehat{\bh}_\text{hybrid} = \widehat{\bx} + \cB(\widehat{\bX})$. 

\subsection{Parameter Estimation}
We can extract estimates of the far-field angles $\left\{\widehat{\theta}_{k_\text{f}}\right\}_{k_\text{f} = 1}^{K_\text{f}}$, near-field angles $\left\{\widehat{\theta}_{k_\text{n}}\right\}_{k_\text{n} = 1}^{K_\text{n}}$, and near-field ranges $\left\{\widehat{r}_{k_\text{n}}\right\}_{k_\text{n} = 1}^{K_\text{n}}$ from the SDP solution as well. \\
{\bf Angle parameter estimation:} The estimates of the far-field and near-field angle parameters $\{\widehat{\phi}_{k_\text{f}}\}_{k_\text{f} = 1}^{K_\text{f}}$,  $\{\widehat{\phi}_{k_\text{n}}\}_{k_\text{n} = 1}^{K_\text{n}}$ are extracted by applying the Vandermonde decomposition to the Toeplitz matrices $\text{Toep}(\widehat{\bu}_{\text{far}})$ and $\text{Toep}(\widehat{\bu}_{\text{near}})$ \cite{VerYang}. 
The estimates of the original angle parameters are then computed as $\widehat{\theta}_{k_\text{f}} = \arcsin \frac{\widehat{\phi}_{k_\text{f}} \lambda_c}{d}$ and $\widehat{\theta}_{k_\text{n}} = \arcsin \frac{\widehat{\phi}_{k_\text{n}} \lambda_c}{d}$, respectively. \\
{\bf Near-field range parameter estimation:} Substituting $\widehat{\bX}$ and $\{\widehat{\phi}_{k_\text{n}}\}_{k_\text{n} = 1}^{K_\text{n}}$ into the identity $\bX^{\star} = \bZ_{\alpha}\bD_{\phi}^H$, where $\bD_{\phi} = \begin{bmatrix}\bd(\phi_1), \cdots,   \bd(\phi_{K_\text{n}})\end{bmatrix}$ and $\bZ_{\alpha} = \begin{bmatrix}  \alpha_1 \bz_1, \cdots, \alpha_{K_\text{n}} \bz_{K_\text{n}}\end{bmatrix}$, we obtain the estimate $\widehat{\bZ}_{\alpha} = \widehat{\bX} (\bD_{\widehat{\phi}}^H)^{\dagger }$, where $(\cdot)^{\dagger }$ denotes the Moore-Penrose inverse. Note that $\bz_{k_\text{n}}$ can only be recovered up to a scaling ambiguity, due to the multiplicative form of $\alpha_{k_\text{n}}\bz_{k_\text{n}}$. To obtain an estimate of $r_{k_\text{n}}$, we develop a heuristic to eliminate the effect of the scaling ambiguity. Let $\widehat{\bg}_{k_\text{n}} = \bB \widehat{\bz}_{k_\text{n}}$, $\bg_{k_\text{n}}^{(1)} = \begin{bmatrix} \left[\bg_{k_\text{n}}\right]_0, \cdots,  \left[\bg_{k_\text{n}}\right]_{N-2}\end{bmatrix}^T$, and $\bg_{k_\text{n}}^{(2)} = \begin{bmatrix} \left[\bg_{k_\text{n}}\right]_1, \cdots,  \left[\bg_{k_\text{n}}\right]_{N-1}\end{bmatrix}^T$. We construct
$\bg_{k_\text{n}}^{0} =  \bg_{k_\text{n}}^{(2)}  \oslash \bg_{k_\text{n}}^{(1)} = \begin{bmatrix} e^{j2\pi \psi_{k_\text{n}}}, e^{j2\pi 3\psi_{k_\text{n}}}, \cdots, e^{j2\pi (2N-3)\psi_{k_\text{n}}}\end{bmatrix}^T$,
where $\oslash$ denotes the element-wise division. Denoting the constant vector $\bn_0 = \begin{bmatrix}1, 3, \cdots, 2N-3 \end{bmatrix}^T$, the estimate of the range $r_{k_\text{n}}$ is then given by 
$
\widehat{r}_{k_\text{n}} = \frac{-j \pi d^2 \cos^2(\widehat{\theta}_{k_\text{n}})}{\lambda_c \bn_0^{\dagger} \log(\widehat{\bg}^0_{k_\text{n}} )}
$.

\section{Numerical Simulations}
\begin{figure}[htb]
   \begin{subfigure}[b]{.49\columnwidth}
    \includegraphics[width=\columnwidth, height=1.3in]{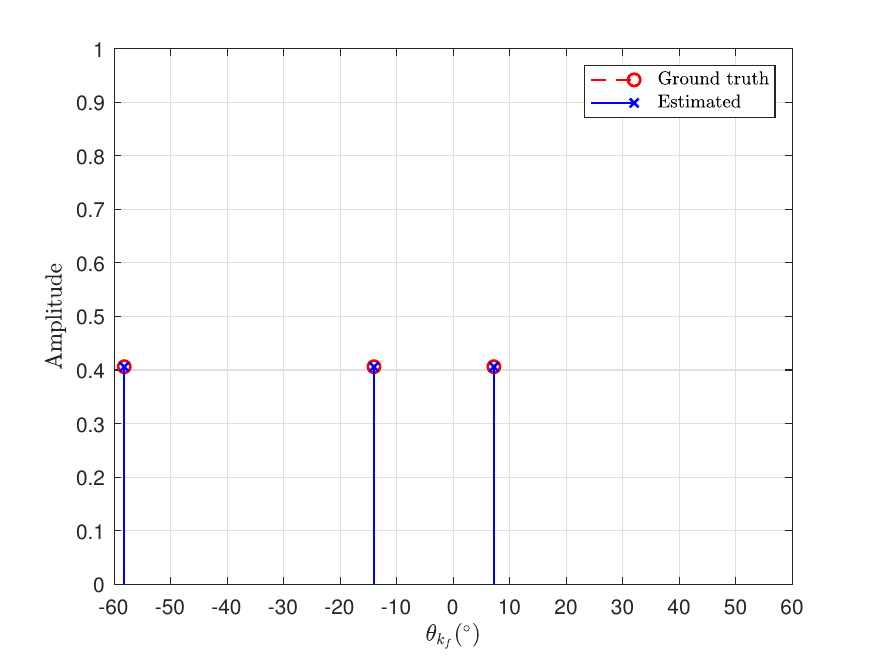}
     \caption{}
   \end{subfigure} 
   \begin{subfigure}[b]{.49\columnwidth}
    \includegraphics[width=\columnwidth, height=1.3in]{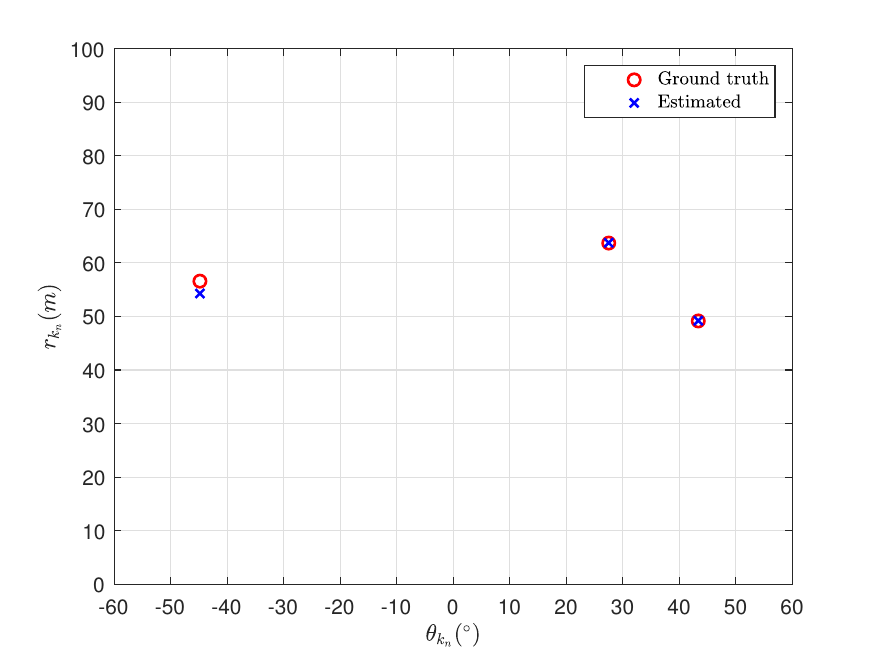}
    \caption{}
   \end{subfigure} 
   \vspace{-0.15cm}
    \caption{The recovered angles and ranges vs ground truth. (a) Estimated far-field angles versus ground truth; (b) estimated near-field angles and ranges versus ground truth. }
    \label{fig:0}
   \end{figure}

We consider a MIMO system with the following configuration: a BS with $N = 256$ antennas, a carrier frequency $f_c = 30$GHz, $N_\text{RF} = 4$ RF chains, and a Rayleigh distance of approximately $327.68$m. A dictionary is formed with columns $\bg(\psi)$, where $\psi$ is discretized to a fine resolution, and a low-rank approximation scheme with $r = 10$ is employed to construct the subspace matrix $\bB$ in (\ref{eq:subspace}). In all experiments, for each $K$, propagation paths are evenly distributed between far-field and near-field regions. In the first illustrative example, we randomly generate a hybrid-field channel with $K = 6$ paths. Figure \ref{fig:0} demonstrates the recovered angle and range parameters by solving (\ref{eq:sdp}), superimposed on the ground truth. It is observed that the angle parameters are recovered with very high accuracy, while the ranges incur some error due to the approximations of the near-field steering vectors and subspace modeling.
   
\begin{figure}[htb]
   \begin{subfigure}[b]{.49\columnwidth}
    \includegraphics[width=\columnwidth, height=1.3in]{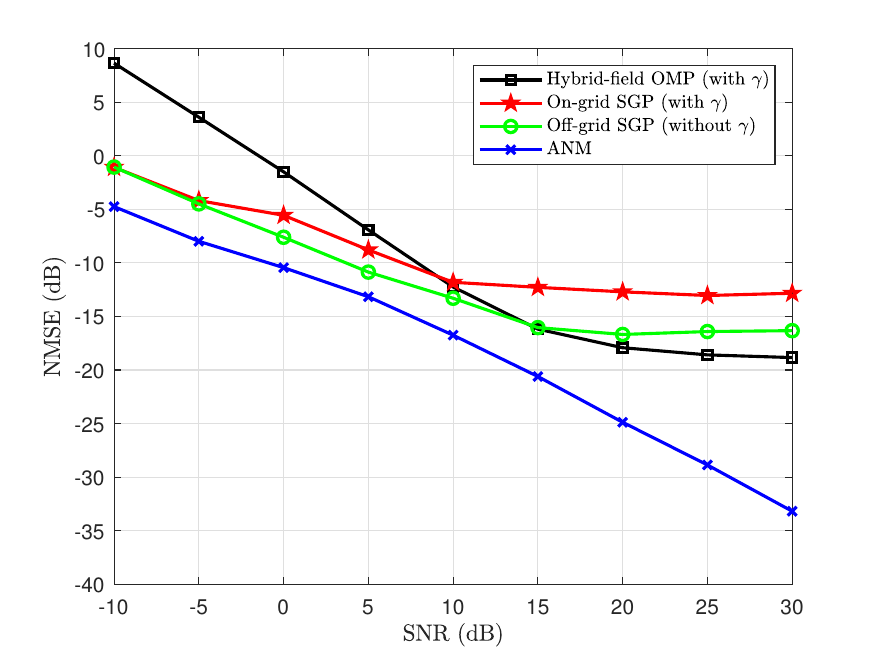}
     \caption{}
   \end{subfigure} 
   \begin{subfigure}[b]{.49\columnwidth}
    \includegraphics[width=\columnwidth, height=1.3in]{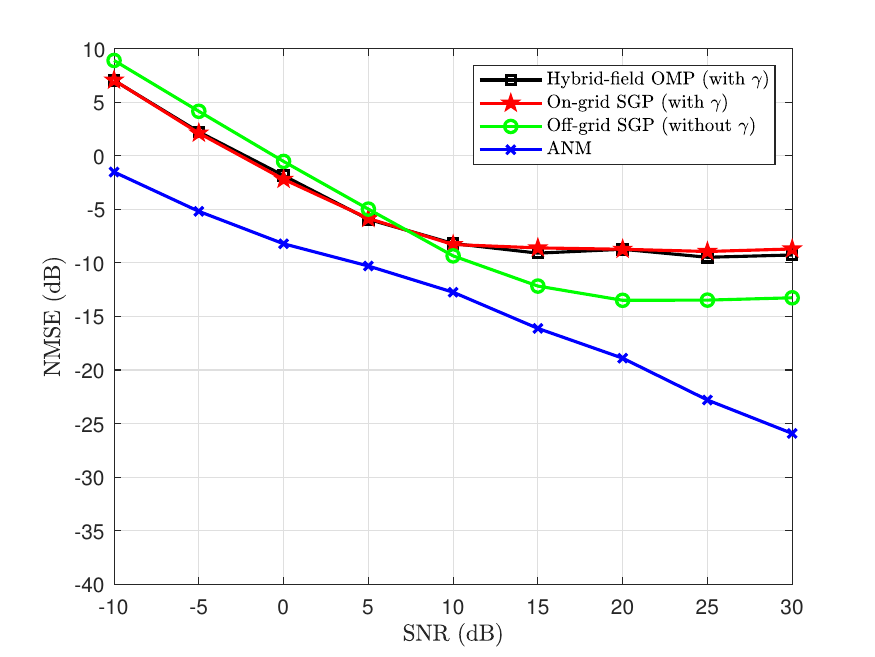}
    \caption{}
   \end{subfigure} 
      \vspace{-0.15cm}
    \caption{The NMSE performance of various hybrid-field channel estimation schemes with respect to different SNRs. (a) No downsampling; (b) downsampling by a factor of $2$.}
    \label{fig:1}
   \end{figure}

Next, we compare the proposed ANM with hybrid-field orthogonal matching pursuit (OMP) \cite{Wei-CL2022} and stochastic gradient pursuit (SGP) \cite{Lei-TSP2024} for channel estimation across various signal-to-noise ratio (SNR) levels. We set $K = 10$ and compare two sampling schemes: i) no downsample, where the number of measurements in (\ref{eqn:y}) equals $N$; ii) downsampling by a factor of $2$. 
%
The normalized mean squared error (NMSE), defined as  $\mathbb{E} \left\{ \| \widehat{\bh} - \bh_{\text{hybrid}}\|_2^2 / \| \bh_{\text{hybrid}}\|_2^2\right\}$, serves as the performance metric. For each SNR level, NMSE is averaged over $50$ simulation trials. Figure \ref{fig:1} shows that the proposed ANM  achieves significantly lower NMSE compared to OMP and SGP. Note that, for OMP and SGP, $\gamma$ is the percentage of the far-field paths, i.e., $\gamma = K_\text{f}/K$. 

\begin{figure}[htb]
   \begin{subfigure}[b]{.49\columnwidth}
    \includegraphics[width=\columnwidth, height=1.3in]{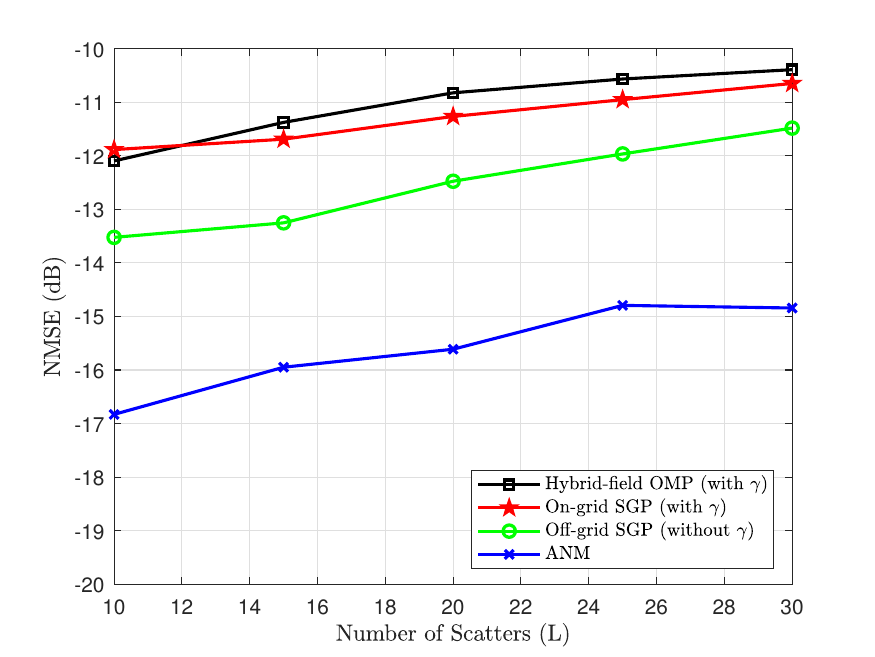}
     \caption{}
   \end{subfigure} 
   \begin{subfigure}[b]{.49\columnwidth}
    \includegraphics[width=\columnwidth, height=1.3in]{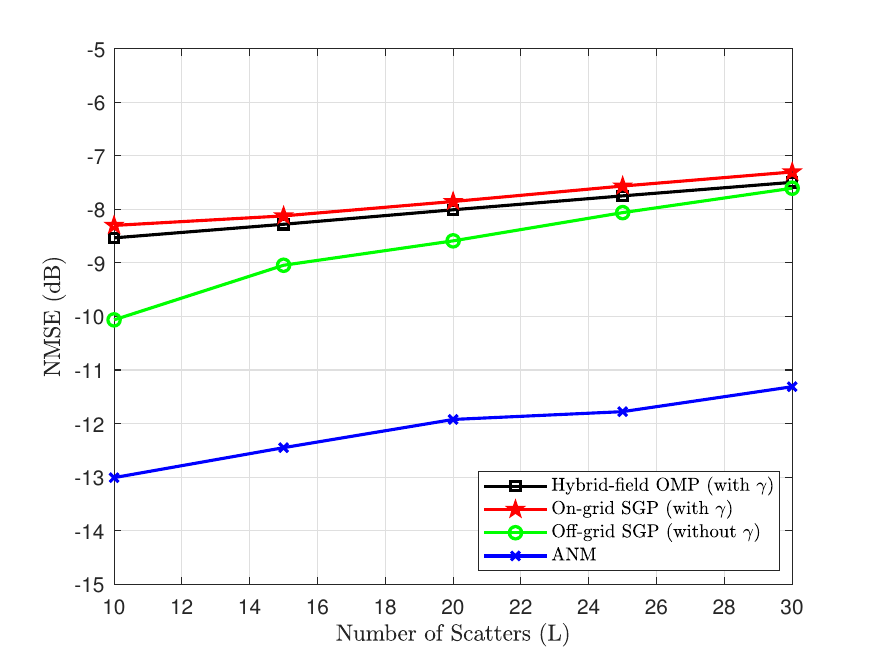}
    \caption{}
   \end{subfigure} 
      \vspace{-0.15cm}
\caption{The NMSE performance of various hybrid-field channel estimation schemes with respect to different $K$. (a) No downsampling; (b) downsampling by a factor of $2$.}
    \label{fig:2}
   \end{figure}

In the third experiment, we investigate the impact of the total number of propagation paths $K$ on channel estimation performance. Figure \ref{fig:2} illustrates the NMSE performance for varying $K$ at a fixed SNR of $10$dB. As expected, channel estimation becomes more challenging as $K$ increases. Nevertheless, the proposed ANM consistently outperforms hybrid-field OMP and SGP. 

\vspace{-0.5cm}
%
%
%
\vspace{-0.3cm}
\section{Conclusion}
In this paper, we proposed a novel approach for hybrid-field channel estimation by leveraging ANM. 
By solving a constrained convex optimization with two ANMs that promote sparsity for the far-field and near-field components, respectively, we accurately estimate the channel and its associated angle and range parameters without requiring discretization. 
A theoretical analysis of the proposed convex demixing scheme and its dual problem is currently under investigation. 

\vfill\pagebreak

%
%

\bibliographystyle{IEEEbib}
\bibliography{strings,hybridfield,refs_DehuiYang}

\end{document}